# On the magnetic phase diagram of a sigma-phase $Fe_{47}Re_{53}$ compound in the H-T plane

Stanisław M. Dubiel[*]

AGH University of Science and Technology, Faculty of Physics and

Applied Computer Science, al. A. Mickiewicza 30, 30-059 Kraków, Poland

M. I. Tsindlekht  and  I. Felner,

Racah Institute of Physics, The Hebrew University,

Jerusalem, Israel 91904

**Abstract**

Sigma-phase $Fe_{47}Re_{53}$ compound was investigated by means of in-field DC and AC magnetization measurements. Field cooled, $M_{FC}$, and zero-field cooled, $M_{ZFC}$, DC magnetization curves were recorded in the magnetic field, H, up to 1000 Oe. AC magnetization measurements were performed at the constant frequency of 1465 Hz in the field up to H=500 Oe. Both types of measurements gave evidence in favor of a re-entrant character of magnetism in the studied sample. Based on characteristic temperatures determined from the measurements magnetic phase diagrams have been outlined in the H-T plane.  The diagrams are similar but not identical. The main difference is that in the diagram constructed based on the DC data there is an intermediate antiferromagnetic-like phase which is absent in the AC diagram. Furthermore, the border lines (irreversibility, cross-over and Curie temperature) follow the power law but with correspondingly different exponents. The studied sample was also characterized by determining the effect of H on the degree of irreversibility and that of frustration.

[*] Corresponding author: Stanislaw.Dubiel@fis.agh.edu.pl

# 1. Introduction

The sigma ($\sigma$) phase is likely the best-known representative example of a family of Frank-Kasper (FK) phases, also known as topologically close-paced (TCP) structures. Its archetype is $\sigma$ in the Fe-Cr system discovered in 1927 [1] and crystallographically identified some 25 years later [2]. Its tetragonal unit cell accommodates 30 atoms distributed over 5 lattice sites having high coordination numbers (12-15). It occurs in alloys in which at least one element is a transition metal. Among binary alloys there are 43 cases known in which the occurrence of $\sigma$ has been revealed [3]. An interest in $\sigma$ has been two-fold: (a) industrial and (b) scientific. The former stems from the fact that it often precipitates in technologically important materials like, for example, ferritic and/or martensitic steels casing a serious deterioration of their useful properties like corrosion resistance, creep strength, impact toughness or tensile ductility. Consequently, its presence in these materials is highly undesired. On the other hand, efforts has been recently undertaken to profit from its high hardness in order to strengthen materials properties e. g. [4]. Scientifically, $\sigma$ has been of interest *per se* due to their complex structure which is further complicated by a lack of stoichiometry. In fact, for a given alloy, $\sigma$ exists in a certain range of composition. Consequently, its physical properties can be tailored by changing the composition. In particular, the magnetic properties turned out to be very sensitive to a chemical composition. For instance in the $Fe_{100-x}V_x$ system the Curie temperature can be changed between few Kelvin and over 300 K [5]. Concerning the magnetism of $\sigma$, it has been so far revealed in four binary Fe-based alloys viz. Fe-V [6], Fe-Cr [7], Fe-Re [8] and Fe-Mo [9]. The magnetic ordering of $\sigma$ in Fe-Cr and Fe-V alloys was originally regarded as ferromagnetic (FM), yet it was recently shown to be more complex than initially anticipated as in both cases it has been evidenced to have a re-entrant character [11,12]. Similar character of magnetism has $\sigma$ in Fe-Re and Fe-Mo alloys [8,9].

The study described in this paper was aimed at shedding more light on the magnetism of $\sigma$ in Fe-Re. According to the previous study performed on $\sigma$-$Fe_{100}Re_x$ (x=43-53) alloys the Curie temperature, $T_C$, ranges between ~65 K for x=43 and ~23K for x=53 [8]. At lower temperatures a transition into a spin-glass (SG) state was detected. The SG state was found to be magnetically heterogeneous i.e. a weak and a strong irreversibility ranges were identified. A transition from FM to SG in re-entrant

SGs takes place at the so-called irreversibility temperature, $T_{ir}$, while a second transition occurs at the so-called cross-over temperature, $T_{co}$. According to the mean-field theory these transitions also occur in an external magnetic field, H. The *loci* of $T_C$, $T_{ir}$ and $T_{co}$ in the H-T coordinates form the so-called H-T magnetic phase diagram. Its construction for a $\sigma$-$Fe_{47}Re_{53}$ sample has been performed in the present study based on in-field DC and AC magnetizations measurements and it is presented in this paper.

## 2. Experimental

### 2.1. The sample

The preparation of the sample of $\sigma$-$Fe_{47}Re_{53}$ was as follows: powders of elemental iron (3N+ purity) and rhenium (4N purity) were mixed in appropriate proportions and masses (2 g), and next pressed to a pellet. The pellet was subsequently melted in an arc furnace under protective atmosphere of argon. The produced ingot was re-melted three times to increase the homogeneity. Afterwards, the ingot was vacuum annealed at 1803 K for 5 hours and, finally, quenched into liquid nitrogen. The mass los of the fabricated sample was below 0.01% of its initial value, therefore the nominal composition has been taken as the real one. X-ray diffraction patterns recorded at room temperature on the powdered sample gave evidence that it had the tetragonal crystallographic structure. More details on structural and electronic properties of this sample are given elsewhere [13].

### 2.2. DC and AC magnetization measurements

DC Magnetization (M) measurements at various applied magnetic fields, H, up to 1000Oe, and in the temperature interval 5 K < T < 60 K have been performed using the commercial (Quantum Design) superconducting quantum interference device (SQUID) magnetometer with sample mounted in gel-caps. Prior to recording the zero-field-cooled (ZFC) curves, the SQUID magnetometer was always adjusted to be in a *"true"* H = 0 state. The temperature dependence of the field-cooled (FC) and the ZFC branches were taken via warming the samples. The real ($\chi'$) and imaginary ($\chi''$) AC magnetic susceptibilities at H up to 500 Oe were measured with a home-made pickup coil method at an AC field amplitude of $h_0$=0.04 Oe at frequencies of 1465 Hz. The AC measurements were performed by using the same SQUID magnetometer.

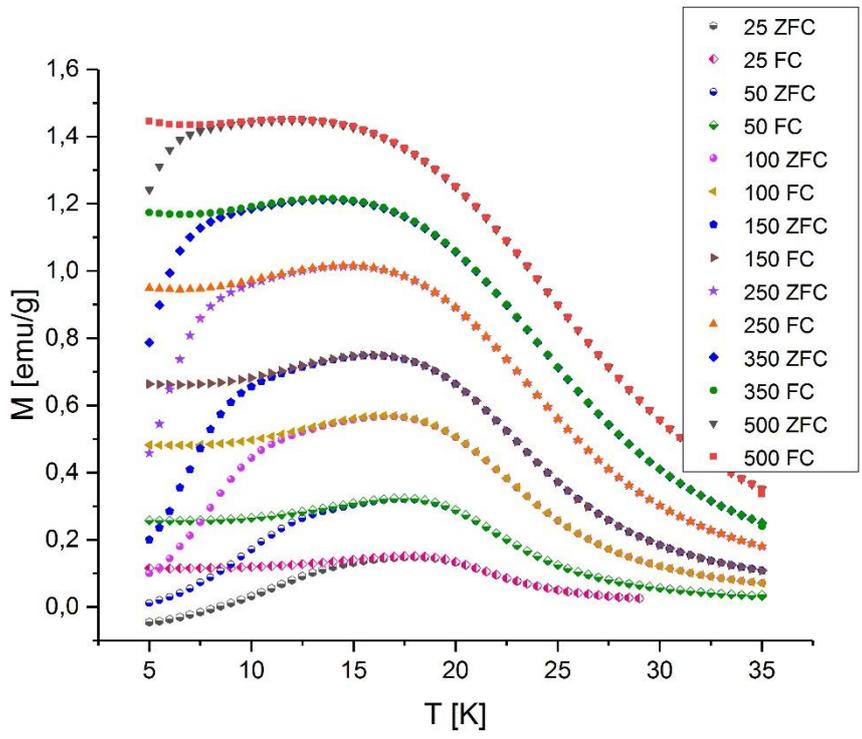

Fig. 1 Selected DC magnetization, M, curves measured vs. temperature, T, in zero-field (ZFC) and in field cooling (FC). Values of the external magnetic field in Oe are given in the legend.

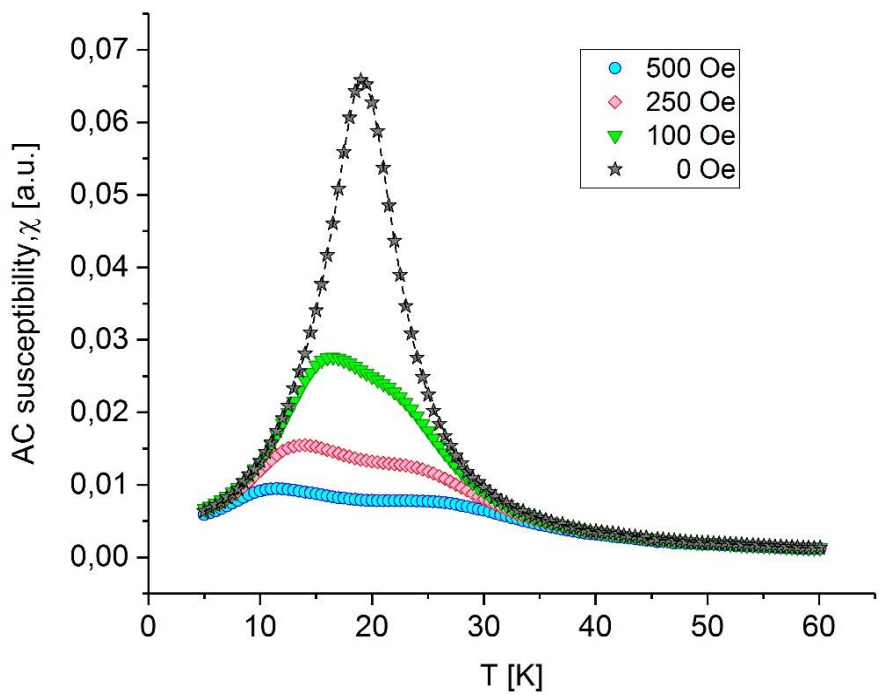

Fig. 2a Real part of the AC magnetic susceptibility versus temperature, T, measured in external magnetic field of different intensity up to 500 Oe. The AC frequency was 1465 Hz and the amplitude of the ac field 0.04 Oe.

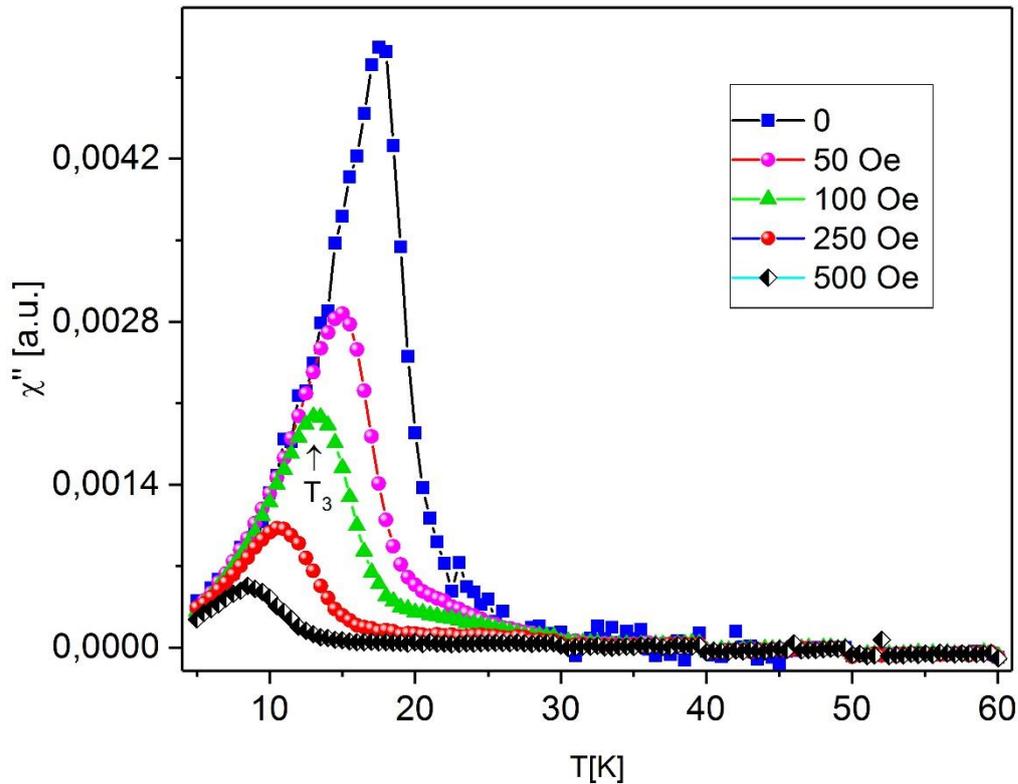

Fig. 2b Imaginary part of the AC magnetic susceptibility, $\chi''$, versus temperature, T, measured in external magnetic field of different intensity as shown in the legend. The AC frequency was 1465 Hz and the amplitude of the field 0.04 Oe.

## 2. Results and discussion

### 2.1. DC measurements

#### 2.1.1. Characteristic temperatures

The temperatures behavior of the $M_{FC}$ and $M_{ZFC}$ curves can be described in terms of three characteristic temperatures: $T_C$ defined by the inflection point, $T_{ir}$ defined by the bifurcation point, and $T_m$ defined by the maximum value of $M_{ZFC}$. $T_C$ can be precisely determined by considering the temperature behavior of $dM_{ZFC}/dT$ – see Fig. 3.

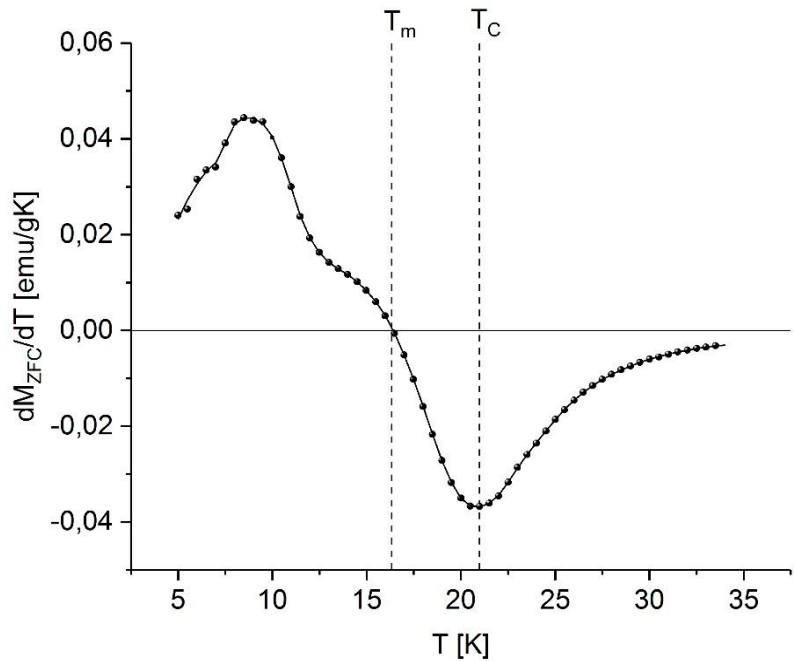

Fig. 3 Temperature behavior of $dM_{ZFC}/dT$ curve for 100 Oe. $T_C$ and $T_m$ positions are marked by vertical dashed lines.

$T_C$ in the present case can be regarded as the Curie temperature which follows from a positive value of the Curie-Weiss temperature, $\Theta_{CW}$. Indeed, as presented in Fig. 4 for the measurement at H=25 Oe, the reciprocal DC susceptibility, $1/\chi$, can be well-fitted to the Curie-Weiss law in the paramagnetic state (PM) yielding $\Theta_{CW}$=21.4 K. The figure gives evidence in favor of effective ferromagnetic interactions in the PM state.

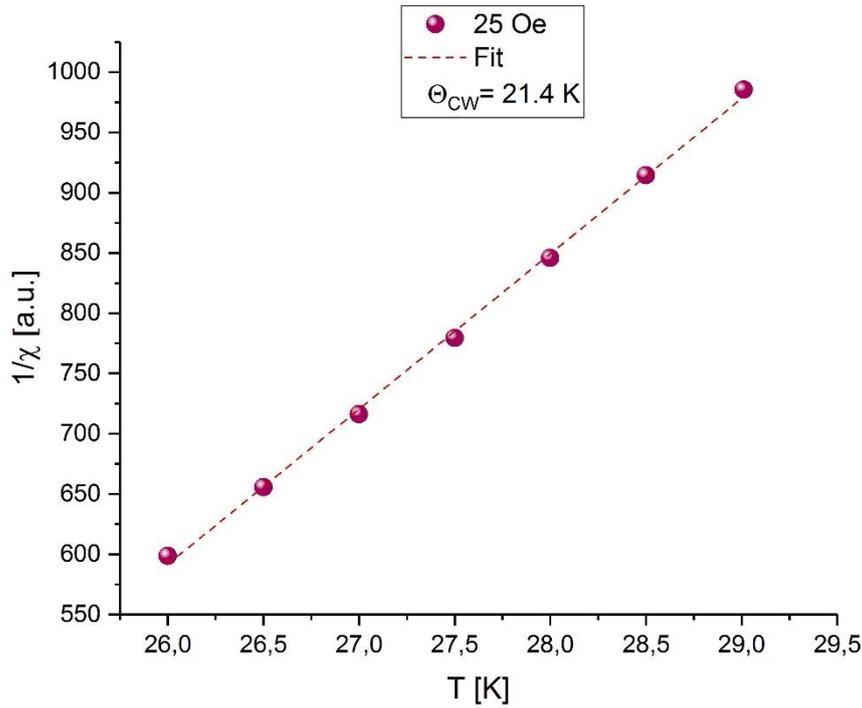

Fig. 4 Reciprocal DC magnetization susceptibility, $1/\chi$, vs. temperature, T, in the paramagnetic state of the investigated sample. The straight line stays for the best-fit to the data in terms of the Curie-Weiss law.

The temperature of bifurcation or irreversibility, $T_{ir}$, can be interpreted as the one indicating a transition from a FM into a SG state. It can be determined considering the difference between $M_{FC}$ and $M_{ZFC}$, $\Delta M$. A set of the $\Delta M$ curves obtained for the studied sample is illustrated in Fig. 5. The temperature at which $\Delta M \geq 0$ is equal by definition to $T_{ir}$.

The third temperature to be considered is the one at which the $M_{ZFC}$ curve has a maximum, $T_m$. Its value can be readily determined considering the temperature behavior of $dM_{ZFC}/dT$ – see Fig. 3. This temperature can be regarded as the one indicating a transition from a weak into a strong irreversibility state of SG, hence it is also known as the cross-over temperature ($T_{co}$).

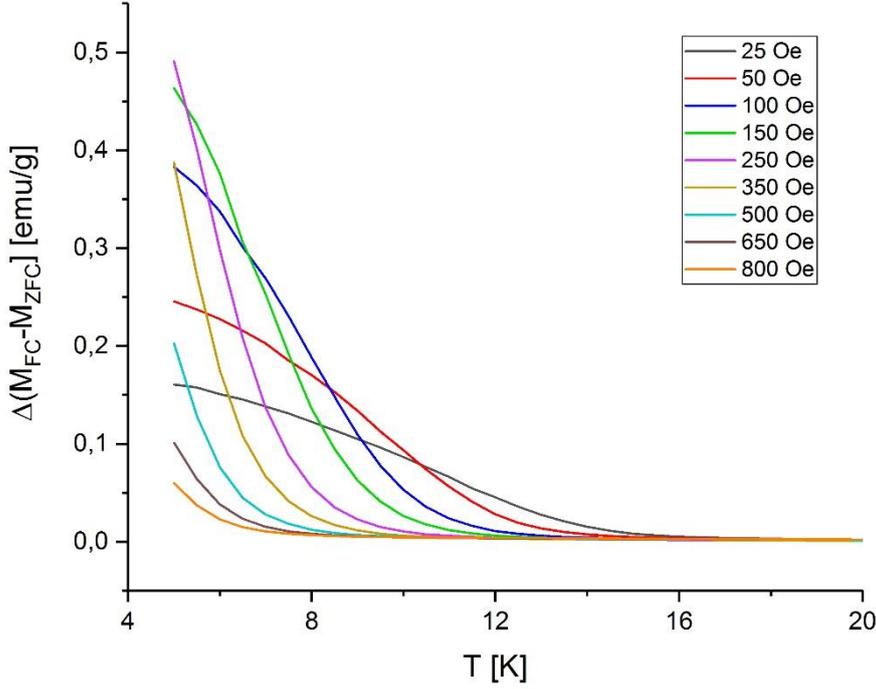

Fig. 5 Difference in the FC and ZFC magnetization curves, ∆M, versus temperature for various values of the external magnetic field as displayed. Based on these curves the irreversibility temperature, $T_{ir}$, was determined.

## 2.1.2. Degree of irreversibility

The M(T)-curves shown in Fig. 1 give a clear evidence that a degree of irreversibility of the SG state depends both on temperature as well as on the strength of the magnetic field. It is also evident that with growing H the difference between the $M_{FC}$ and $M_{ZFC}$ curves gets smaller. The latter can be obviously related to a decrease of a spin canting caused by the applied magnetic field. In order to quantitatively describe this effect we calculated a difference in area under the $M_{FC}(T,H)$ and $M_{ZFC}(T,H)$ curves, ∆M (H):

$$\Delta M(H) = \int_{T_1}^{T_2} (M_{FC}(H) - M_{ZFC}(H)) dT \qquad (1)$$

Where $T_1$=5 K and $T_2$=$T_{ir}$.

Such obtained quantity, ΔM, is presented in Fig. 6. It is clear that the degree of irreversibility decreases sharply with H up to around 500 Oe. The ΔM(H)-dependence can be well-described by an exponential function and the best-fit in terms of such function is marked by a dashed line.

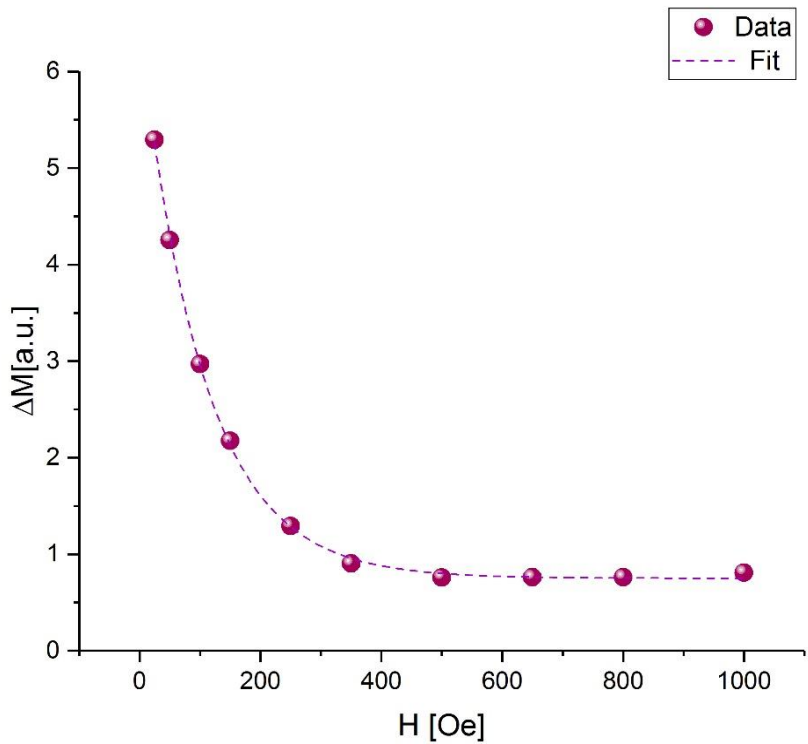

Fig. 6 Integrated difference of $M_{FC}$ and $M_{ZFC}$ magnetization curves, ΔM, vs. magnetic field, H. The dashed line shows the best exponential fit to the data.

Indeed, the $M_{FC}$ curve does not saturate. Instead it slightly increases at low T, probably due to some small magnetic particles not detectable by X-ray diffraction.

**2.1.2. Degree of frustration**

As a measure of a degree of a spin frustration, DF, the ratio between the Curie temperature, $T_C$, and the Curie-Weiss one, $\Theta_{CW}$, was suggested and used [13]. Intuitively, one would expect that DF decreases with H. The present study permits verification of this expectation. The effect of H on both quantities relevant for determination of DF is shown in Fig. 7. Whereas $T_C$ steadily increases with H, the dependence of $\Theta_{CW}$ can be divided into two ranges: (I) for H ≤ 100 Oe, where an increase is observed, and (II) for H > 100 Oe where $\Theta_{CW}$ decreases. Based on the results shown in Fig. 7, the dependence of FD on H has been found and presented in Fig. 8.

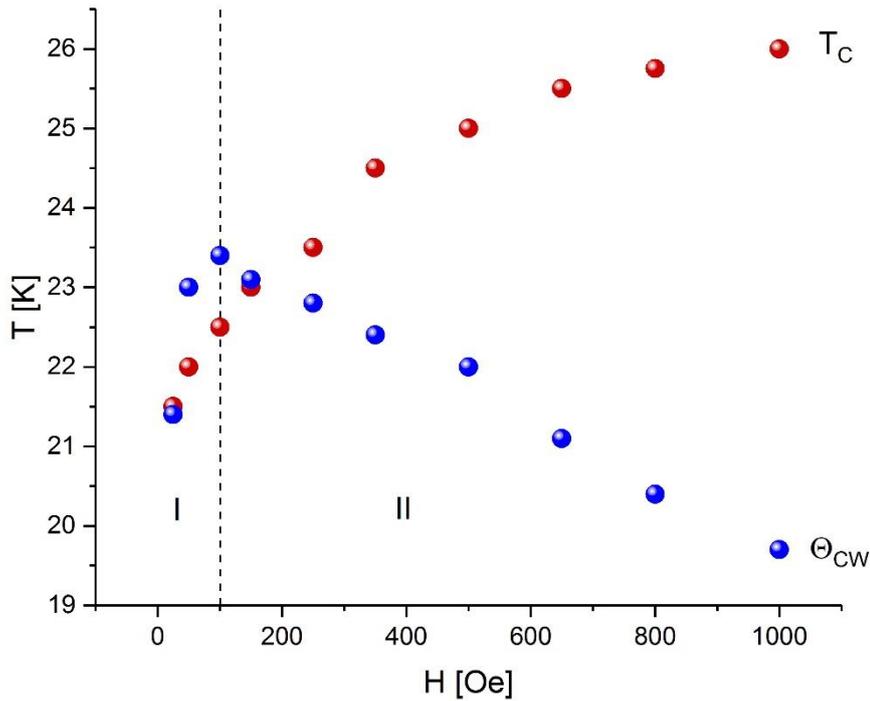

Fig. 7 Dependence of the Curie temperature, $T_C$, and that of the Curie-Weiss temperature, $\Theta_{CW}$, on the applied magnetic field, H. The dashed vertical line indicates a division of the $\Theta_{CW}$-dependence into two ranges: (I) where an increase of $\Theta_{CW}$ takes place, and (II) where a decrease of $\Theta_{CW}$ occurs.

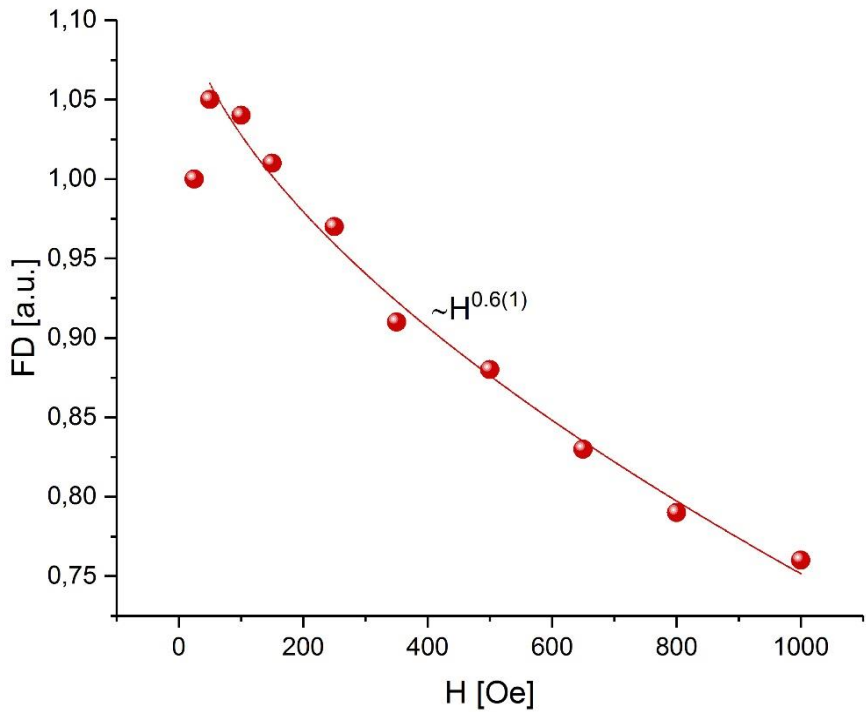

Fig. 8 The degree of frustration, DF, vs. magnetic field, H. The full line is the best fit to the data for H ≥ 50 Oe in terms of a power law.

Evidently, FD increases only in a narrow range of H viz. 25-50 Oe. For higher H-values a decrease is observed and its H-dependence can be well-described in terms of a power law with an exponent 0.6(1).

**2.1.3. Magnetic phase diagram in the H-T plane**

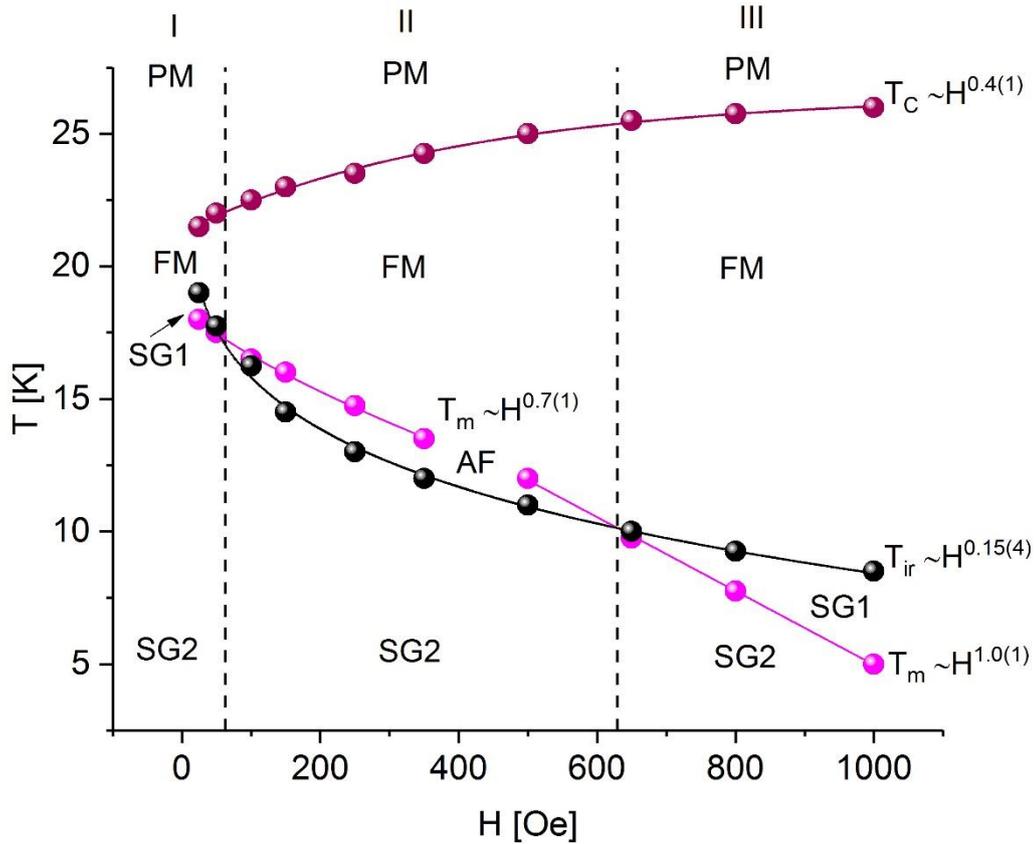

Fig. 9 Magnetic phase diagram of the sigma-phase $Fe_{47}Re_{53}$ compound as deduced from the DC in-field magnetization measurements. The meaning of the symbols and acronyms is given in the text.

Based on the revealed magnetic field dependences of the three temperatures i.e. $T_C$, $T_{ir}$ and $T_m$ a magnetic phase diagram for the investigated sample in the H-T plane has been constructed and displayed in Fig. 9. Three ranges can be distinguished: I, II and III. In the range I, which exists for $0 \leq H \leq 50$ Oe, an expected re-entrant transition viz. PM→FM→SG takes place. Here $T_{ir} < T_m$ and the SG state can be divided into the one with a weak irreversibility (designated as SG1) and the other with a strong irreversibility (denoted as SG2). In the range II, that exists for $50 \leq H \leq 630$ Oe, $T_{ir} < T_m$. Also the $M_{FC}$ curve shows a concave shape for $T \leq T_m$ what can be understood in terms of an effective antiferromagnetic coupling between magnetic moments ascribed to Fe atoms sited on various sites. Further support for such interpretation comes from theoretical calculations [8], according to which magnetic moments on sites A, C and D can be coupled antiferromagnetically. Thus in this range the sequence of a transition is as follows: PM→FM→AF→SG. Obviously, the effective coupling between the moments depends on the strength of the applied magnetic field. Consequently, we see a change of this coupling with H. Finally, in the range III, the ferromagnetic coupling prevails, and consequently we see again a standard i.e. PM→FM→SG re-entrant transition. The measured $T_C(H)$, $T_{ir}(H)$ and $T_m(H)$ dependences can be next used for validation of different predictions relevant to the issue. Thus according to the mean-field theory (MFT), the T-H relationship reads as follows:

$$T(H) = T(0) - a \cdot H^{\varphi} \qquad (1)$$

Where T can be either $T_{ir}$ or $T_m$.

Three different predictions concerning the value of $\varphi$ and thus the T-H relationships can be found throughout the literature as far as the re-entrant SGs are concerned: (1) $\varphi = 2/3$ (for $T_{ir}$), (2) $\varphi = 2$ (for $T_m$) [14], and (3) $\varphi = 1$ [15].

All three sets of the data have been successfully fitted to Eq.(1). The best-fits have been indicated by solid lines in Fig. 8. The $T_{ir}$ data could have been correctly reproduced with only one value of $\varphi=0.15$ in the whole H-range, which is much less than the expected value of 2/3. In turn, the position of the maximum, $T_m$, follows initially i.e. up to H=350 Oe the power law with $\varphi=0.7(1)$, while in the range of $500 \leq H \leq 1000$ Oe the value of $\varphi=1$. In other words one observes a crossover in the field dependence of $T_m$ which takes place for H between 350 and 500 Oe. Noteworthy, the crossover from a non-linear to a linear behavior in $T_m$ was also observed for another FK-phase, namely the Laves $NbFe_2$ compound [16], whereas for $\sigma$-$Fe_{100-x}V_x$ samples with x=33.5 and x=34.1 such behavior was revealed both in $T_{ir}$ as well as in $T_m$ [12]. Finally, it should be noted that also the Curie temperature shows a power law dependence on H, yet in this case it grows with the field as $H^{0.4}$. An increase of $T_C$ with H was also reported for the $NbFe_2$ [16] and for the $\sigma$-FeV [12]. All these compounds exhibit an itinerant character of magnetism.

## 2.2. AC results

### 2.2.1. Characteristic temperatures

Five characteristic temperatures, $T_1$, $T_2$, $T_3$, $T_4$ and $T_5$ can be deduced from the AC susceptibility measurements. Namely, $T_1$, $T_2$ as position of the peaks in the real part of the AC curves (Fig. 10), $T_3$ the position of the peak in the imaginary part of the susceptibility, $T_4$ and $T_5$ positions of the minima in the temperature derivative of the real part of the susceptibility – see the inset in Fig. 10. It has turned out that $T_2 \approx T_5$ and their values are close to the $T_C$-values found from the DC measurements, hence they can be regarded as the magnetic ordering (Curie) temperature. Also very similar to each other are $T_3$ and $T_4$ temperatures (they are marked in Fig. 11 by diamond and pentagon, respectively). In other words the AC measurements yielded three different temperatures, $T_2=T_I$, $T_1=T_{II}$ and $T_3=T_{III}$, and they have been used to construct the H-T phase diagram as shown in Fig. 11.

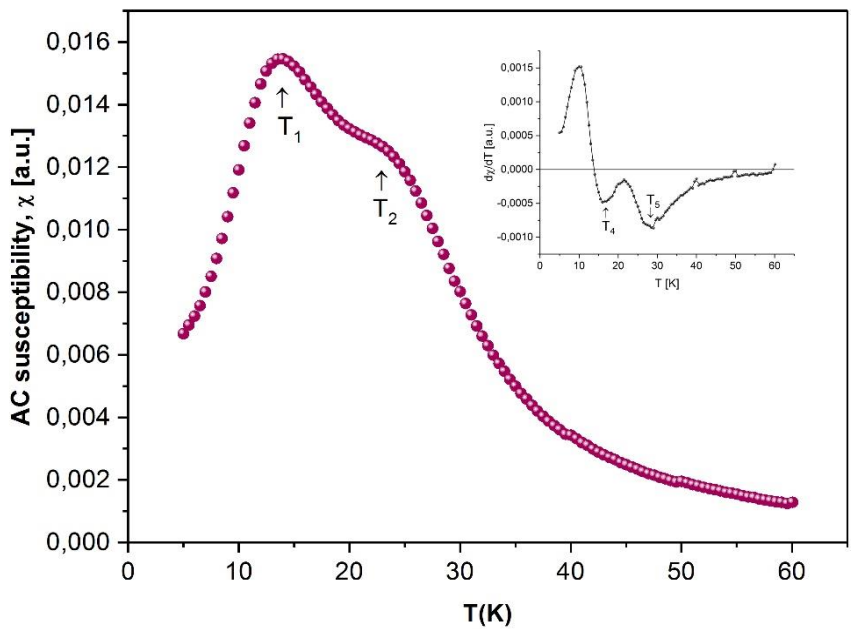

Fig. 10 Real part of the AC susceptibility, $\chi$, vs. temperature, T, as measured in an external magnetic field of 250 Oe. The inset shows $d\chi/dT$ with $T_4$ and $T_5$ indicated by arrows.

### 2.2.2. Magnetic phase diagram in the H-T plane.

The H-T magnetic phase obtained from the in-field AC magnetization measurements has both similarities as well as differences in comparison with the one deduced based on the in-field DC magnetization measurements. Concerning the former, the re-entrant character of magnetism of the investigated sample is clear. The lines characteristic of such phenomenon: (1) irreversibility line marking a transition from an intermediate FM-like state into a SG-state with a weak irreversibility (denoted here as SG1), and (2) cross-over line indicating a border between the SG1 and the ground state SG state with a strong irreversibility (denoted here as SG2) can be drawn as a *loci* of the $T_{II}(H)$ and $T_{III}(H)$, respectively. Both of them follow the MFT prediction i.e. they decrease with H [14], but only the irreversibility line does it quantitatively i.e. with 0.7 that is practically equal to the predicted $\varphi \approx 2/3$.

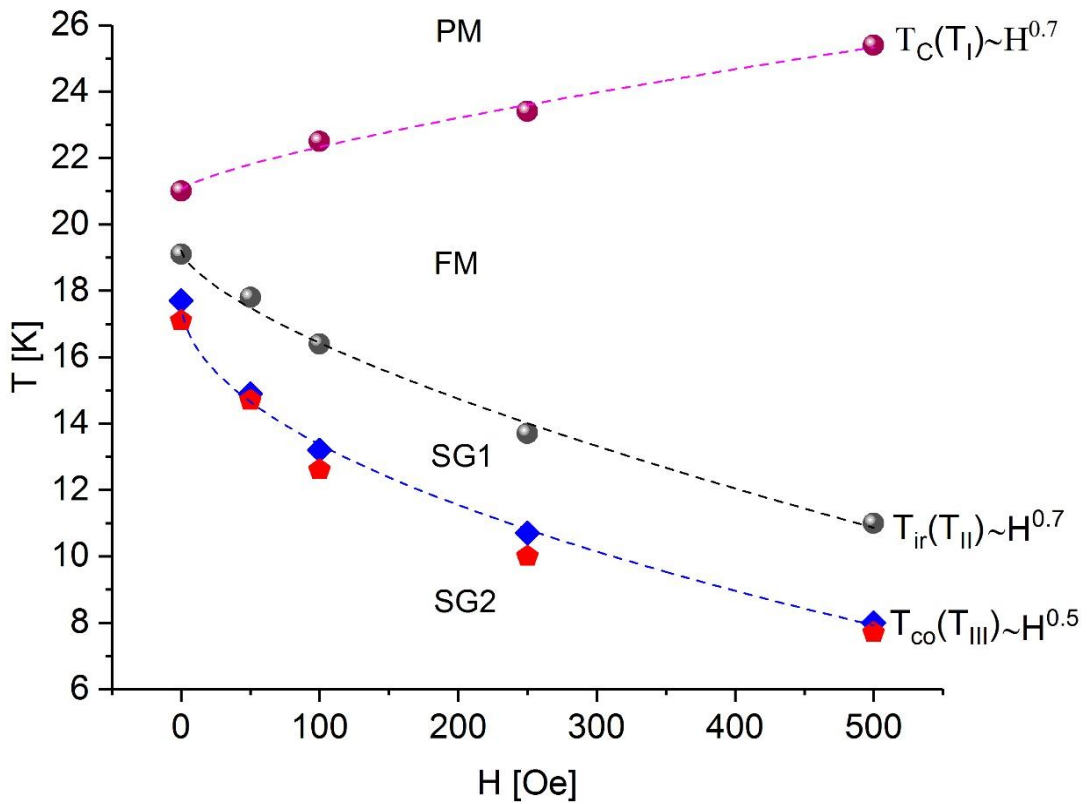

Fig. 11 Magnetic phase diagram of the sigma-phase $Fe_{47}Re_{53}$ compound as deduced from the AC in-field magnetization measurements. The meaning of the symbols and acronyms is given in the text.

Regarding the differences, first no intermediate AF-like phase has been detected from the AC data, second the irreversibility line does not show a cross-over behavior, and third the power law exponents for all three lines are different. Concerning the differences, it is worth noticing that they were recently reported for the $NbFe_2$ compound [16]. Consequently, one has to keep in mind when validating theoretical models of SGs that experimental results can significantly depend on the method used for their determination, hence the validation may not be unique.

## 3. Summary

Performed DC and AC magnetization measurements enabled construction of the magnetic phase diagrams of the sigma-phase $Fe_{47}Re_{53}$ compound in the H-T coordinates. The obtained phase diagrams show both similarities and differences. Concerning the similarities, both of them give a clear evidence that the magnetism of the studied compound has a re-entrant character. However, they differ concerning the existence of an intermediate AF-like phase that exists following the DC data. Three lines in the H-T plane characteristic of the re-entrant spin-glasses viz. (1) the border line between the paramagnetic (PM) and the ferromagnetic (FM) states, (2) the border line between the FM and the spin glass state with a weak irreversibility (SG1), and (3) the border line between SG1 and the spin-glass state with a strong irreversibility (SG2) follow the power law but with correspondingly different exponents. Furthermore, the line (2) determined from the DC data shows a cross-over for $350 \leq H \leq 500$ Oe from a nonlinear to a linear dependence on H. In general, values of the exponents for the lines (2) and (3) do not agree with those predicted by the mean-field model [14]. The exception is the line (2) determined from the AC measurements which is in line with that prediction. The DC data permitted also determination of the effect of H on the degree of irreversibility and that of frustration.

## Acknowledgements

This work was financed by the Faculty of Physics and Applied Computer Science AGH UST statutory tasks within subsidy of Ministry of Science and Higher Education in Warsaw.